# Wireless Local Area Networks with Multiple-Packet Reception Capability

***Abstract*** **-** Thanks to its simplicity and cost efficiency, wireless local area network (WLAN) enjoys unique advantages in providing high-speed and low-cost wireless services in hot spots and indoor environments. Traditional WLAN medium-access-control (MAC) protocols assume that only one station can transmit at a time: simultaneous transmissions of more than one station causes the destruction of all packets involved. By exploiting recent advances in PHY-layer multiuser detection (MUD) techniques, it is possible for a receiver to receive multiple packets simultaneously. This paper argues that such multipacket reception (MPR) capability can greatly enhance the capacity of future WLANs. In addition, it provides the MAC-layer and PHY-layer designs needed to achieve the improved capacity. First, to demonstrate MUD/MPR as a powerful capacity-enhancement technique, we prove a "super-linearity" result, which states that the system throughput per unit cost increases as the MPR capability increases. Second, we show that the commonly deployed *binary* exponential backoff (BEB) algorithm in today's WLAN MAC may not be optimal in an MPR system, and that the optimal backoff factor increases with the MPR capability: the number of packets that can be received simultaneously. Third, based on the above insights, we design a joint MAC-PHY layer protocol for an IEEE 802.11-like WLAN that incorporates advanced PHY-layer blind detection and MUD techniques to implement MPR.

Corresponding Author: Ying Jun (Angela) Zhang
Email: yjzhang@ie.cuhk.edu.hk
Tel:/Fax: (852) 2609 8465

## I. INTRODUCTION

### *A. Motivation*

The last decade has witnessed a surge of interest in wireless local area networks (WLAN), where mobile stations share a common wireless medium through contention-based medium access control (MAC). In WLANs, collision of packets occurs when more than one station transmits at the same time, causing a waste of bandwidth. The recent advances in multiuser detection (MUD) techniques [1] open up new opportunities for resolving collisions in the physical (PHY) layer. For example, in CDMA [2][2] or multiple-antenna [3] systems, multiple packets can be received simultaneously using MUD techniques without collisions. WLAN capacity can be expected to improve greatly with such techniques. To date, most previous work on MUD has been restricted to the PHY layer. To fully utilize the multipacket reception (MPR) capability for capacity enhancement in WLAN, however, it is essential to understand the fundamental impact of MPR on the MAC-layer design. With this backdrop, this paper is a first study of the MAC-layer throughput performance and the collision resolution schemes for WLANs with MPR.

### *B. Key Contributions*

The key contributions of this paper are as follows:

- To demonstrate MUD/MPR as a powerful capacity-enhancement technique at the system level, we analyze the MAC-layer throughput of WLANs with MPR capability under both finite-node and infinite-node assumptions. In contrast to previous work in [4]-[7], our model is sufficiently general to cover both carrier-sensing and non-carrier-sensing networks. We prove that in random-access WLANs, network throughput increases *super-linearly* with the MPR capability of the channel. That is, throughput divided by $M$ increases as $M$ increases, where $M$ is the number of packets that can be resolved simultaneously. The super-linear throughput scaling implies that the achievable throughput *per unit cost* increases with MPR capability of the channel. This provides a strong incentive to deploy MPR in next-generation wireless networks.
- We study the effect of MPR on the MAC-layer collision resolution scheme. When packets collide, an exponential backoff (EB) scheme is commonly used to schedule the retransmissions, in which the waiting time of the next retransmission will get multiplicatively longer for each collision incurred. In the commonly adopted binary exponential backoff (BEB) scheme (e.g., used in Ethernet [16], WiFi [17], etc.), the multiplicative (a backoff factor) is equal to 2. We show in this paper that BEB does not necessarily yield the close-to-optimal network throughput when the wireless channel can accommodate more than one simultaneous packet transmission. Instead, the optimal backoff factor heavily depends on the relative durations of idle, collision, and success slots. BEB is far from optimum for both non-carrier-sensing networks and carrier-sensing networks operated in basic access mode. The optimal backoff factor increases with the MPR capability. Meanwhile, BEB is close to optimum for carrier-sensing networks when RTS/CTS access mode is adopted.

- Built on the theoretical underpinnings established above, we propose a practical protocol to fully exploit the MPR capability in IEEE 802.11-like WLANs. In contrast to [8]-[9], we consider not only the MAC layer protocol design, but also the PHY-layer signal processing to enable MPR in distributed random-access WLANs. As a result, the proposed protocol can be implemented in a fully distributed manner with marginal modification of current IEEE 802.11 MAC.

### *C. Related Work on MPR and Collision Resolution Schemes*

The first attempt to model a general MPR channel in random-access wireless networks was made by Ghez, Verdú, and Schwartz in [4]-[5] in 1988 an 1989, respectively. The stability properties of slotted ALOHA systems with MPR were studied under a simple infinite-user and single-buffer assumption. In 2005, Naware et al studied the stability and delay of finite-user ALOHA systems over both symmetric and asymmetric MPR channels [6]. In [7], the authors studied the stability and capacity regions of MPR networks when an optimal MAC protocol is available. The previous analyses have mainly focused on the stability of networks with MPR. Moreover, they are limited to either a specific MAC protocol, namely ALOHA [4]-[6], or an idealized optimal MAC [7]. Little work has been done to investigate the effect of MPR on the performance of practical random-access wireless networks with backoff schemes. Our paper here is an attempt along this direction. In addition, the super-linearity result as a motivating factor for MPR in this paper is also new.

Protocols that exploit the MPR capability of networks have been studied by Zhao and Tong in [8]-[9]. In [8], a multi-queue service room (MQSR) MAC protocol was proposed for networks with heterogeneous users. The drawback of the MQSR protocol is its high computational cost due to updates of the joint distribution of all users' states. To reduce complexity, a suboptimal dynamic queue protocol was proposed in [9]. In both protocols, access to the common wireless channel is controlled by a central controller, which grants access to the channel to an appropriate subset of users at the beginning of each slot. By contrast, this paper provides a distributed IEEE 802.11-like MAC protocol that is more applicable to practical implementations.

Exponential Backoff (EB) as a collision resolution technique has been extensively studied in different contexts [11]-[14]. Stability upper bound of Binary EB (BEB) has been given by Goodman under a finite-node model in [11] and recently improved by Al-Ammal in [12]. The throughput and delay characteristics of a slightly modified EB scheme have been studied in [13] in the context of slotted ALOHA. The characteristics of EB in steady state is further investigated in [14] in time slotted wireless networks with equal slot length. All the existing work on EB has assumed that the wireless channel can only accommodate one ongoing transmission at a time. This paper is a first attempt to look at EB for an MPR system.

The remainder of this paper is organized as follows. In Section II, we describe the system model and introduce the background knowledge on MUD and EB. In Section III, we prove that the maximum achievable throughput of MPR WLAN scales super-linearly with the MPR capability of the channel. In Section IV, we investigate the effect of MPR on EB. We illustrate that the widely used BEB scheme is no longer close-to-optimal in MPR networks. In Section V, a MAC-PHY protocol is proposed to realize MPR in IEEE 802.11 WLANs. Section VI concludes this paper.

## II. SYSTEM MODEL

### *A. System Description*

We consider a fully connected random-access network with $N$ infinitely backlogged mobile stations. We assume that the time axis is divided into slots and packet transmissions start only at the beginning of a slot. In addition, after each transmission, the transmitting stations have a means to discover the result of the transmission, i.e., success or failure. If the transmission fails due to collision, the colliding stations will schedule retransmissions according to a collision resolution scheme (e.g., EB). We assume that the channel has the capability to accommodate up to $M$ simultaneous transmissions. In other words, packets can be received correctly whenever the number of simultaneous transmissions is no larger than $M$. When more than $M$ stations contend for the channel at the same time, collision occurs and no packet can be decoded. We refer to $M$ as MPR capability hereafter.

In our model, the length of a time slot is not necessarily fixed and may vary under different contexts [10]. We refer to this variable-length slot as *backoff slot* hereafter. In WLANs, the length of a backoff slot depends on the contention outcome (hereafter referred to as channel status). Let $T_i$ denote the length of an idle time slot when nobody transmits; $T_c$ denote the length of a collision time slot when more than *M* stations contend for the channel; and $T_s$ denote the length of a time slot due to successful transmission when the number of transmitting stations is anywhere from 1 to *M*. The durations of $T_i$, $T_c$, and $T_s$ depend on the underlying WLAN configuration. For non-carrier-sensing networks such as slotted ALOHA, the stations are not aware of the channel status and the duration of all backoff slots are equal to the transmission time of a packet. That is,

$$T_{slot} = T_i = T_c = T_s = L/R, \tag{1}$$

where *L* is the packet size and *R* is the data transmission rate of a station. On the other hand, for carrier-sensing networks, stations can distinguish between various types of channel status and the durations of different types of slots may not be the same. For example, in IEEE 802.11 DCF basic access mode,

$$\begin{cases} T_i = \sigma \\ T_s = H + L/R + SIFS + \delta + ACK + DIFS + \delta \\ T_c = H + L/R + DIFS + \delta \end{cases} \tag{2}$$

where $\sigma$ is the time needed for a station to detect the packet transmission from any other station and is typically much smaller than $T_c$ and $T_s$; *H* is the transmission time of PHY header and MAC header; *ACK* is the transmission time of an ACK packet; $\delta$ is the propagation delay; and *SIFS* and *DIFS* are the inter-frame space durations [17]. Similarly, in IEEE 802.11 DCF request-to-send/clear-to-send (RTS/CTS) access scheme, the slot durations are given by

$$\begin{cases} T_i = \sigma \\ T_s = RTS + SIFS + \delta + CTS + SIFS + \delta \\ \qquad + H + L/R + SIFS + \delta + ACK + DIFS + \delta \\ T_c = RTS + DIFS + \delta \end{cases} \tag{3}$$

where *RTS* and *CTS* denote the transmission time of RTS and CTS packets, respectively. By allowing the durations of $T_i$, $T_c$, and $T_s$ to vary according to the underlying system, the analysis of this paper applies to a wide spectrum of various WLANs, including both non-carrier-sensing and carrier-sensing networks.

*B. Multiuser Detection*

This subsection briefly introduces the PHY layer MUD techniques used to decode multiple packets at the receiver. Let $x_k(n)$ denote the data symbol transmitted by user *k* in symbol duration *n*. If there are *K* stations transmitting together, then the received signal at a receiver is given by

$$\mathbf{y}(n)=\sum_{k=1}^{K}\mathbf{h}_k(n)x_k(n)+\mathbf{w}(n)=\mathbf{H}(n)\mathbf{x}(n)+\mathbf{w}(n)\,. \qquad (4)$$

where $\mathbf{w}(n)$ denotes the additive noise, $\mathbf{H}(n)=\left[\mathbf{h}_1(n),\mathbf{h}_2(n),\cdots\mathbf{h}_K(n)\right]$, and $\mathbf{x}(n)=\left[x_1(n),\cdots,x_K(n)\right]^T$. In CDMA systems, vector $\mathbf{h}_k$ is the spreading sequence of user *k*. In multiple antenna systems, $\mathbf{h}_k$ is the channel vector, with the $m^{\text{th}}$ element being the channel coefficient from user *k* to the $m^{\text{th}}$ receive antenna.

The receiver attempts to obtain an estimate of the transmitted symbols $\mathbf{x}(n)$ from the received vector $\mathbf{y}(n)$. To this end, various MUD techniques have been proposed in the literature. For example, the zero-forcing (ZF) receiver is one of the most popular linear detectors. It multiplies the received vector by the pseudo-inverse of matrix $\mathbf{H}(n)$, denoted by $\mathbf{H}^{+}(n)$, and the decision statistics become

$$\mathbf{r}^{ZF}(n)=\mathbf{H}^{+}(n)\mathbf{y}(n)=\mathbf{x}(n)+\mathbf{H}^{+}(n)\mathbf{w}(n)\,. \qquad (5)$$

The minimum-mean-square-error (MMSE) receiver is the optimal linear detector in the sense of maximizing the signal-to-interference-and-noise ratio (SINR). The decision statistics is calculated as

$$\mathbf{r}^{MMSE}(n)=\left(\mathbf{H}(n)\mathbf{H}^{H}(n)+\eta\mathbf{I}\right)^{-1}\mathbf{H}^{H}(n)\mathbf{y}(n) \qquad (6)$$

where $\mathbf{I}$ is the identity matrix, and $\eta$ is the variance of the additive noise. Given the decision statistics, an estimate of $x_k(n)$ can be obtained by feeding the $k^{\text{th}}$ element of $\mathbf{r}^{ZF}(n)$ or $\mathbf{r}^{MMSE}(n)$ into a quantizer.

Other MUD techniques include maximum-likelihood (ML), parallel interference cancellation (PIC), successive interference cancellation (SIC), etc. Interested readers are referred to [1] for more details.

*C. Exponential Backoff*

EB adaptively tunes the transmission probability of a station according to the traffic intensity of the network. It works as follows. A backlogged station sets its backoff timer by randomly choosing an integer within the range $\left[0,W-1\right]$, where *W* denote the size of the contention window. The backoff timer is decreased by one following each backoff slot. The station transmits a packet in its queue once the backoff timer reaches zero. At the first transmission attempt of a packet, $W=W_0$, referred to as the minimum contention window. Each time the transmission is unsuccessful, the $W$ is multiplied by a backoff factor $r$. That is, the contention window size $W_i = r^i W_0$ after $i$ successive transmission failures.

## III. Super-Linear Throughput Scaling in WLANs with MPR

This section investigates the impact of MPR on the throughput of random-access WLANs. In particular, we prove that the maximum achievable throughput scales super-linearly with the MPR capability $M$. In practical systems, $M$ is directly related to the cost (e.g., bandwidth in CDMA systems or antenna in multi-antenna systems). Super-linear scaling of throughput implies that the achievable throughput *per unit cost* increases with $M$. This provides a strong incentive to consider MPR in next-generation wireless networks.

As mentioned earlier, the transmission of stations is dictated by the underlying EB scheme. To capture the fundamentally achievable spectrum efficiency of the system, the following analysis assumes that each station transmits with probability $p_t$ in an arbitrary slot, without caring how $p_t$ is achieved. The assumption will be made more rigorous in Section IV, which relates $p_t$ to EB parameters such as $r$ and $W_0$.

### *A. Throughput of WLANs with MPR*

Define throughput to be the average number of information bits transmitted successfully per second. Let $S_N(M, p_t)$ denote the throughput of a WLAN with $N$ stations when each station transmits at probability $p_t$ and the MPR capability is $M$. Then, $S_N(M, p_t)$ can be calculated as the ratio between the average payload information bits transmitted per backoff slot to the average length of a backoff slot as follows.

$$\begin{aligned} S_N(M, p_t) &= \frac{E[\text{payload information bits transmitted in a backoff slot}]}{E[\text{length of a backoff slot}]} \\ &= \frac{\sum_{k=1}^{M} k \Pr\{X = k\} L}{P_{idle} T_i + P_{coll} T_c + P_{succ} T_s} \quad \text{bits/sec} \end{aligned} \tag{7}$$

In the above, $X$ is a random variable denoting the number attempts in a slot.

$$\Pr\{X = k\} = \binom{N}{k} p_t^k (1 - p_t)^{N-k}. \tag{8}$$

Let $$P_{idle} = \left(1 - p_t\right)^N \tag{9}$$

be the probability that a backoff slot is idle;

$$P_{succ} = \sum_{k=1}^{M} \Pr\left\{X = k\right\} = \sum_{k=1}^{M} \binom{N}{k} p_t^k \left(1 - p_t\right)^{N-k} \tag{10}$$

be the probability that a backoff slot is busy due to successful packet transmissions; and

$$P_{coll} = \sum_{k=M+1}^{N} \Pr\left\{X = k\right\} = \sum_{k=M+1}^{N} \binom{N}{k} p_t^k \left(1 - p_t\right)^{N-k} \tag{11}$$

be the probability that a backoff slot is busy due to collision of packets.

The throughput of non-carrier-sensing networks such as slotted ALOHA can be obtained by substituting (1) into (7), which leads to following expression.

$$S_N\left(M, p_t\right) = \frac{\sum_{k=1}^{M} k \Pr\left\{X = k\right\} L}{T_{slot}} = R \sum_{k=1}^{M} k \binom{N}{k} p_t^k \left(1 - p_t\right)^{N-k} \ \text{bits/sec}. \tag{12}$$

Similarly, the throughput of carrier-sensing networks, such as IEEE 802.11 DCF basic-access mode and RTS/CTS access mode, can be obtained by substituting (2) and (3) into (7) respectively.

We now derive the asymptotic throughput when the population size $N$ approaches infinity. In this case, we assume that (i) the system has a non-zero asymptotic throughput; and (ii) the number of attempts in a backoff slot is approximated by a Poisson distribution with an average attempt rate $\lambda = Np_t$. Both of these assumptions are valid under an appropriate EB scheme, which will be elaborated in Section IV. Let $S_\infty\left(M, \lambda\right)$ be the asymptotic throughput when MPR capability is $M$ and average attempt rate is $\lambda$. Then, we derive from (7) that

$$\begin{aligned} S_\infty\left(M, \lambda\right) = \lim_{N \to \infty} S_N &= \frac{L}{P_{idle}T_i + P_{coll}T_c + P_{succ}T_s} \sum_{k=1}^{M} k \Pr\left\{X = k\right\} \\ &\overset{(a)}{=} \frac{L}{P_{idle}T_i + P_{coll}T_c + P_{succ}T_s} \sum_{k=1}^{M} k \frac{\lambda^k}{k!} e^{-\lambda} \\ &= \frac{L\lambda}{P_{idle}T_i + P_{coll}T_c + P_{succ}T_s} \sum_{k=0}^{M-1} \frac{\lambda^k}{k!} e^{-\lambda} \\ &= \frac{L\lambda}{P_{idle}T_i + P_{coll}T_c + P_{succ}T_s} \Pr\left\{X \le M - 1\right\} \ \text{bits/sec} \end{aligned} \tag{13}$$

where Equality (a) is due to the Poisson approximation. In particular, when $T_{slot} = T_i = T_c = T_s = L/R$,

$$S_\infty(M,\lambda) = R\sum_{k=0}^{M-1}\frac{\lambda^{k+1}}{k!}e^{-\lambda} = R\lambda\Pr\{X \le M-1\} \text{ bits/sec} \tag{14}$$

*B. Super-Linear Throughput Scaling*

Having derived the throughput expressions for both finite-population and infinite-population models, we now address the question: how does throughput scale as *M* increases. In particular, we are interested in the behavior of the *maximum* throughput when the channel has a MPR capability of *M*. This directly relates to the channel-access efficiency that is achievable in MPR networks.

Given *M*, the maximum throughput can be achieved by optimizing the transmission probability $p_t$ (or equivalently $\lambda$ in the infinite-population model). The optimal transmission probability can in turn be obtained by adjusting the backoff factor *r* in practical WLANs, as will be discussed in Section IV. Let $S_N^*(M) = S_N\left(M, p_t^*(M)\right)$ and $S_\infty^*(M) = S_\infty\left(M, \lambda^*(M)\right)$ denote the maximum achievable throughputs, where $p_t^*(M)$ and $\lambda^*(M)$ denote the optimal $p_t$ and $\lambda$ when the MPR capability is *M*, respectively. In Theorem 1, we prove that the throughput scales super-linearly with *M* in non-carrier-sensing network with infinite population. In other words, $S_\infty^*(M)/M$ is an increasing function of *M*. In Theorem 2, we further prove that $S_\infty^*(M)/MR$ approaches 1 when $M \to \infty$. This implies that the throughput penalty due to distributed random access diminishes when *M* is very large. In Theorem 3 in Appendix A, we prove that the same super-linearity holds for WLANs with finite population.

***Theorem 1*** (*Super-linearity*)*:* $S_\infty^*(M)/M$ is an increasing function of *M*.

It is obvious that at the optimal $\lambda^*(M)$

$$\left.\frac{\partial S_\infty(M,\lambda)}{\partial\lambda}\right|_{\lambda=\lambda^*(M)} = R\sum_{k=0}^{M-1}\frac{(k+1)\left(\lambda^*(M)\right)^k}{k!}e^{-\lambda^*(M)} - R\sum_{k=0}^{M-1}\frac{\left(\lambda^*(M)\right)^{k+1}}{k!}e^{-\lambda^*(M)} = 0. \tag{15}$$

Consequently,

$$\sum_{k=0}^{M-1}\frac{\left(\lambda^*(M)\right)^k}{k!}e^{-\lambda^*(M)} = \frac{\left(\lambda^*(M)\right)^M}{(M-1)!}e^{-\lambda^*(M)}, \tag{16}$$

or
$$\Pr\{X \le M-1\}\Big|_{\lambda=\lambda^*(M)} = M\Pr\{X = M\}\Big|_{\lambda=\lambda^*(M)}. \tag{17}$$

To prove Theorem 1, we show that $S_\infty^*(M+1)/(M+1) \ge S_\infty^*(M)/M$ for all *M* in the following.

$$S_\infty^*(M+1) = S_\infty\left(M+1, \lambda^*(M+1)\right) \tag{18}$$

$$\geq S_\infty\left(M+1,\lambda^*(M)\right)=R\sum_{k=0}^{M-1}\frac{\lambda^*(M)^{k+1}}{k!}e^{-\lambda^*(M)}+R\frac{\lambda^*(M)^{M+1}}{M!}e^{-\lambda^*(M)}$$
$$=S_\infty\left(M,\lambda^*(M)\right)+R\lambda^*(M)\Pr\left\{X=M\right\}\Big|_{\lambda=\lambda^*(M)}$$
$$=\frac{M+1}{M}S_\infty^*\left(M\right)$$

where the last equality is due to (14) and (17). Therefore, we have

$$\frac{S_\infty^*(M+1)}{M+1}\geq\frac{S_\infty^*(M)}{M}\quad\forall M. \qquad \square$$

It is obvious that in a WLAN with MPR capability of *M*, the maximum possible throughput is *MR* when there exists a perfect scheduling. In practical random-access WLANs, the actual throughput is always smaller than *MR*, due to the throughput penalty resulting from packet collisions and idle slots. For example, the maximum throughput is well known to be $Re^{-1}$ when $M=1$. Theorem 2 proves that the throughput penalty diminishes as *M* becomes large. That is, the maximum throughput approaches *MR* even though the channel access is based on random contentions.

***Theorem 2*** (*Asymptotic channel-access efficiency*): $\lim_{M\to\infty}\frac{S_\infty^*(M)}{MR}=1$.

Before proving Theorem 2, we present the following two lemmas.

***Lemma 1***: (a) $\lim_{M\to\infty}S_\infty(M)\big/\lambda R=1$ for any attempt rate $\lambda<M$; (b) $\lim_{M\to\infty}S_\infty(M)\big/\lambda R=0$ for any attempt rate $\lambda>M$; and (c) $\lim_{M\to\infty}S_\infty(M)\big/\lambda R=0.5$ for attempt rate $\lambda=M$.

Due to the page limit, only the proof of Lemma 1(a) is presented here, as it is more relevant to the proofs of Lemma 2 and Theorem 2. Interested readers are referred to Appendix B for the proof of Lemma 1(b) and 1(c).

*Proof of Lemma 1*(*a*):

$$S_\infty(M,\lambda)=R\lambda\Pr\left\{X\leq M-1\right\}=R\lambda\sum_{k=0}^{M-1}\frac{\lambda^k}{k!}e^{-\lambda}=R\lambda\left(1-\sum_{k=M}^{\infty}\frac{\lambda^k}{k!}e^{-\lambda}\right)$$
$$\geq R\lambda\left(1-z^{-M}\sum_{k=M}^{\infty}\frac{(\lambda z)^k}{k!}e^{-\lambda}\right)\geq R\lambda\left(1-z^{-M}e^{\lambda(z-1)}\right)\quad\forall z>1 \qquad (19)$$

Let $f(z)=R\lambda\left(1-z^{-M}e^{\lambda(z-1)}\right)$ be the lower bound of $S_\infty(M)$. By solving

$$\frac{\partial f(z)}{\partial z}=R\lambda\left(Mz^{-M-1}e^{\lambda(z-1)}-\lambda z^{-M}e^{\lambda(z-1)}\right)=0, \qquad (20)$$

it can be easily found that $z^* = M/\lambda$ maximizes $f(z)$ and

$$\frac{f(z^*)}{\lambda R} = 1 - \left(\frac{\lambda}{M}\right)^M e^{M\left(1-\frac{\lambda}{M}\right)} \tag{21}$$

Since $z^* > 1$, $\lambda < M$. Let $\lambda = cM$ where $c < 1$. Eqn. (21) can be written as

$$\frac{f(z^*)}{\lambda R} = 1 - \left(ce^{(1-c)}\right)^M \tag{22}$$

It is obvious that $ce^{(1-c)} < 1 \ \ \forall c \neq 1$. (23)

Therefore,

$$\lim_{M\to\infty} \frac{S_\infty(M,\lambda)}{\lambda R} \geq \lim_{M\to\infty} \frac{f^*(z)}{\lambda R} = \lim_{M\to\infty}\left(1 - \left(ce^{(1-c)}\right)^M\right) = 1, \tag{24}$$

On the other hand, the first equality of (19) implies

$$\frac{S_\infty(M,\lambda)}{\lambda R} \leq 1. \tag{25}$$

Combining (24) and (25), we have

$$\lim_{M\to\infty} \frac{S_\infty(M)}{\lambda R} = 1 \ \forall \lambda < M, \tag{26}$$

and Lemma 1(a) follows. □

***Lemma 2***: The optimal attempt rate $\lambda^*(M) < M$ and $\lim_{M\to\infty} \lambda^*(M)\big/M = 1$.

*Proof of Lemma 2:* The mode of Poisson distribution is equal to $\lfloor \lambda \rfloor$, where $\lfloor \bullet \rfloor$ denotes the largest integer that is smaller than or equal to the argument. When $\lambda \geq M$,

$$\Pr\{X = M\} > \Pr\{X = i\} \ \ \forall 0 \leq i \leq M-1, \tag{27}$$

which conflicts with Eqn. (17). Therefore, the optimal attempt rate

$$\lambda^*(M) < M. \tag{28}$$

Combining (12), (17), (28) and Lemma 1, we have

$$\lim_{M\to\infty} M \Pr\{X = M\}\Big|_{\lambda=\lambda^*(M)} = 1. \tag{29}$$

Let $\lambda^* = cM$ where $c < 1$. Eqn. (29) can be written as

$$\lim_{M\to\infty} \frac{(cM)^M}{(M-1)!} e^{-cM} = 1, \tag{30}$$

and $$c = \lim_{M\to\infty} \frac{\left((M-1)!\right)^{\frac{1}{M}}}{M} e^c = \lim_{M\to\infty} \frac{\left(M!\right)^{\frac{1}{M}}}{M} e^c = e^{-(1-c)} \quad (31)$$

where the last equality is due to the Stirling's formula [15].

Solving Eqn. (31), we have

$$\lim_{M\to\infty} \frac{\lambda^*}{M} = \lim_{M\to\infty} c = 1. \qquad \square$$

*Proof of Theorem 2:* From Lemma 1 and Lemma 2, it is obvious that $\lim_{M\to\infty} S_\infty(M) \big/ \lambda^*(M)R = 1$.

The above results are illustrated in Fig. 1, where $S_\infty^*(M)\big/MR$ is plotted as a function of *M* in non-carrier-sensing slotted ALOHA systems. The figure shows that $S_\infty^*(M)\big/MR$ increases with *M*.

In Theorem 1-3, super-linearity is proved assuming the network is non-carrier-sensing. In Fig. 2 and Fig. 3, the optimal throughput $S_\infty^*(M)$ and $S_\infty^*(M)\big/M$ are plotted for carrier-sensing networks, respectively, with system parameters listed in Table I. The figures show that system throughput is greatly enhanced due to the MPR enhancement in the PHY layer. Moreover, the super-linear throughput scalability holds for carrier-sensing networks when *M* is relatively large.

## IV. Impact of MPR on EB in WLAN MAC

In this section, we study the characteristic behavior of WLAN MAC and EB when the channel has MPR capability. We first establish the relationship between transmission probability $p_t$ (or $\lambda$) and EB parameters including backoff factor *r* and minimum contention window $W_0$. Based on the analysis, we will then study how the optimal backoff strategy changes with the MPR capability *M*.

### *A. Transmission Probability*

We use an infinite-state Markov chain, as shown in Fig. 4, to model of operation of EB with no retry limit. The reason for the lack of a retry limit is that it is theoretically more interesting to look at the limiting case when the retry limit is infinitely large. Having said this, we note that the analysis in our paper can be easily extended to the case where there is a retry limit. The state in the Markov chain in Fig. 4 is the backoff stage, which is also equal to the number of retransmissions experienced by the station. As mentioned in Section II, the contention window size is $W_i = r^i W_0$ when a station is in state *i*. In the figure, $p_c$ denotes the conditional collision probability, which is the probability of a collision seen by a packet being transmitted on the channel. Note that $p_c$ depends on the transmission probabilities of stations other than the transmitting one. In our model, $p_c$ is assumed to be independent of the backoff stage of the transmitting station. In our numerical results, we show that the analytical results obtained under this assumption are very accurate when *N* is reasonably large.

With EB, transmission probability $p_t$ is equal to the probability that the backoff timer of a station reaches zero in a slot. Note that the Markov process of MPR networks is similar to the ones in [10, 14], except that the conditional collision probability $p_c$ is different for $M > 1$. Therefore, Eqn. (32) can be derived in a similar way as [10, 14]:

$$p_t = \frac{2(1-rp_c)}{W_0(1-p_c)+1-rp_c} \tag{32}$$

where $rp_c < 1$ is a necessary condition for the steady state to be reachable. The detailed derivation of (32) is omitted due to page limit. Interested readers are referred to [10, 14]. Likewise, the conditional collision probability $p_c$ is equal to the probability that there are *M* or more stations out of the remaining $N-1$ stations contending for the channel. We thus have the following relationship:

$$p_c = 1 - \sum_{k=0}^{M-1} \binom{N-1}{k} p_t^k (1-p_t)^{N-k-1}. \tag{33}$$

It can be easily shown that $p_t$ is a decreasing function of $p_c$ for any $r > 1$ in (32). Meanwhile, $p_c$ is an increasing function of $p_t$ in (33). Therefore, the curves determined by (32) and (33) have a unique intersection corresponding to the root of the nonlinear system. By solving the nonlinear system (32)-(33) numerically for different *N*, we plot the analytical results of $Np_t$ in Fig. 5. In the figures, BEB is adopted. That is, $r = 2$. The minimum contention window size $W_0 = 16$ or 32. To validate the analysis, the simulation results are plotted as markers in the figures. In the simulation, the data are collected by running 5,000,000 rounds after 1,000,000 rounds of warm up.

From the figures, we can see that the analytical results match the simulations very well. Moreover, it shows that $Np_t$ converges to a constant quantity when *N* becomes large. This is a basic assumption in the previous section when we calculated the asymptotic throughput. The constant quantity that $Np_t$ converges to can be calculated as follows.

For large *N*, the number of attempts in a slot can be modeled as a Poisson process. That is,

$$\Pr\{X = k\} = \frac{\lambda^k}{k!} e^{-\lambda} \qquad (34)$$

where
$$\lambda = \lim_{N\to\infty} Np_t . \qquad (35)$$

The conditional collision probability in this limiting case is given by

$$\lim_{N\to\infty} p_c = \Pr\{X \geq M\} = 1 - \sum_{k=0}^{M-1} \frac{\lambda^k}{k!} e^{-\lambda} . \qquad (36)$$

When the system is steady, the total attempt rate $\lambda = \lim_{N\to\infty} Np_t$ should be finite. Therefore,

$$\lim_{N\to\infty} p_t = \lim_{N\to\infty} \frac{2(1-rp_c)}{W_0(1-p_c)+1-rp_c} = 0 , \qquad (37)$$

which implies

$$\lim_{N\to\infty} p_c = \frac{1}{r} . \qquad (38)$$

Combining (36) and (38), we get the following equation

$$\sum_{k=0}^{M-1} \frac{\lambda^k}{k!} e^{-\lambda} = 1 - \frac{1}{r} . \qquad (39)$$

$\lambda$ can then be calculated numerically from (39) given $M$ and $r$. To validate the Poisson assumption, we plot $\lambda$ in Fig. 5. The figure shows that $Np_t$ calculated from (32) and (33), does converge to $\lambda$ when $N$ is large.

Note that the relationship between $p_t$, $\lambda$, and EB established above do not depend on the duration of the underlying backoff slots, and therefore can be applied in both non-carrier-sensing and carrier-sensing networks.

Before leaving this sub-section, we validate another assumption adopted in Section III. That is, EB guarantees a non-zero throughput when $N$ approaches infinity. To this end, the throughput of slotted ALOHA is plotted as a function of $N$ in Fig. 6 when BEB is adopted. It can be seen that the throughputs with the same $M$ converge to the same constant as $N$ increases, regardless of the minimum contention window $W_0$. Similar phenomenon can also be observed in carrier-sensing networks, as illustrated in Fig. 7, where the throughput of IEEE 802.11 WLAN with basic access mode is plotted with detailed system parameters listed in Table I. The asymptotic throughput when $N$ is very large depends only on the MPR capability $M$ and the backoff factor $r$.

*B. Optimal Backoff Factor*

In Section III, we have investigated the maximum network throughput that is achieved by optimal transmission probability $p_t^*(M)$ and $\lambda^*(M)$. The previous sub-section shows that transmission probability is a function of backoff factor *r*. Mathematically, the optimal *r* that maximizes throughput can be obtained by solving the equation $\partial S(M)/\partial r = 0$.

In this section, we investigate how the optimal backoff factor *r* changes with the MPR capability *M*. In Fig. 8 and Fig. 9, we plot the throughput as a function of *r* for both non-carrier sensing networks and carrier-sensing networks in basic-access mode. From the figure, we can see that the optimal *r* that maximizes throughput increases with *M* for moderate to large *M*. This observation can be intuitively explained for non-carrier-sensing networks by (14), (39), and Lemma 1 as follows. Eqns. (14) and (39) indicate that

$$\frac{S_\infty(M,\lambda)}{\lambda R} = \Pr\{X \le M-1\} = 1 - \frac{1}{r}. \tag{40}$$

As Lemma 1 indicates, when *M* is large, $S_\infty(M,\lambda)/\lambda R$ increases with *M* and eventually approaches 1. Consequently, *r* increases with *M*.

As the figures show, the throughput decreases sharply when *r* moves from the optimal $r^*$ to 1. On the other hand, the curve is flat when *r* is larger than the $r^*$. Therefore, in order to avoid dramatic throughput degradation, it is not wise to operate *r* in the region between 1 and $r^*$. Note that when *M* is large, $r^*$ is larger than 2. This implies that the widely used BEB might be far from optimal in MPR WLANs. To further see how well BEB works, we plot the ratio of the throughput obtained by BEB to the maximum achievable throughput in Fig. 10. The optimal *r* that achieves the maximum throughput is plotted versus *M* in Fig. 11. In the figures, we can see that BEB only achieves a small fraction of the maximum achievable throughput when *M* is large in non-carrier-sensing and IEEE 802.11 basic-access mode. For example, when $M = 10$ BEB only achieves about 80 percent of the maximum throughput in non-carrier-sensing networks. In RTS/CTS mode, in contrast, the performance of BEB is close to optimal for a large range of *M*. Therefore, we argue from an engineering point of view that BEB (i.e., $r = 2$) is a good choice for RTS/CTS access scheme, while on the other hand tuning *r* to the optimal is important for non-carrier-sensing and basic-access schemes. Moreover, it is always safe to set *r* to be large enough, based on the observations from Fig. 8 and Fig. 9.

## V. MPR PROTOCOL FOR IEEE 802.11 WLAN

Having investigated the fundamental performance of WLANs with MPR, we now propose a practical protocol to implement MPR in the widely adopted IEEE 802.11 WLAN. In particular, the proposed protocol consists of both the MAC protocol and PHY signal processing scheme. To limit the scope, we focus on RTS/CTS mechanism in this section. A brief discussion on the deployment of MPR in basic-access mode can be found in Appendix C.

Throughout this section, we assume that the MPR capability is brought by the multiple antennas mounted at the access point (AP). This assumption complies with the hardware request of the latest MIMO-based WLAN standards. However, the proposed MAC-PHY protocol can be easily extended to CDMA networks, as the received signal structures in multi-antenna and CDMA systems are almost the same (refer to Section II-C).

### *A. MAC Protocol Description*

The MAC protocol closely follows the IEEE 802.11 RTS/CTS access mechanism, as illustrated in Fig. 12. A station with a packet to transmit first sends an RTS frame to the AP. In our MPR MAC model, when multiple stations transmit RTS frames at the same time, the AP can successfully detect all the RTS frames if and only if the number of RTSs is no larger than $M$. When the number of transmitting stations exceeds $M$, collisions occur and the AP cannot decode any of the RTSs. The stations will retransmit their RTS frames after a backoff time period according to the original IEEE 802.11 protocol. When the AP detects the RTSs successfully, it responds, after a SIFS period, with a CTS frame that grants transmission permissions to all the requesting stations. Then the transmitting stations will start transmitting DATA frames after a SIFS, and the AP will acknowledge the reception of the DATA frames by an ACK frame.

The formats of the RTS and Data frames are the same as those defined in 802.11, while the CTS and ACK frames have been modified to accommodate multiple transmitting stations for MPR. In particular, there are $M$ receiver address fields in the CTS and ACK frames to identify up to $M$ intended recipients.

As described above, our MPR MAC is very similar to the original IEEE 802.11 MAC. In fact, to maintain this similarity in the MAC layer, the challenge is pushed down to the physical layer. For example, in the proposed MPR MAC, the AP is responsible to decode all the RTSs transmitted simultaneously. However, due to the random-access nature WLAN, the AP has no priori knowledge of who the senders are and the channel state information (CSI) on the corresponding links. This imposes a major challenge on the PHY layer, as the MUD techniques introduced in Section II, such as ZF and MMSE cannot be directly applied. To tackle these problems, we introduce the physical layer techniques in next subsection.

*B. PHY-Layer Signal Processing Mechanism*

In this subsection, we propose a PHY mechanism to implement MPR in IEEE 802.11. The basic idea is as follows. RTS packets are typically transmitted at a lower data rate than the data packets in IEEE 802.11. This setting is particularly suitable for blind detection schemes which can separate the RTS packets without knowing the prior knowledge of the senders' identities and CSI [18, 19]. Upon successfully decoding the RTS packets, the AP can then identify the senders of the packets. Training sequences, to be transmitted in the preamble of the data packets, are then allocated to these users to facilitate channel estimation during the data transmission phase. Since the multiple stations transmit their data packets at the same time, their training sequences should be mutually orthogonal. In our system, no more than $M$ simultaneous transmissions are allowed. Therefore, a total of $M$ orthogonal sequences are required to be predefined and made known to all stations in the BSS. The sequence allocation decision is sent to the users via the CTS packet.

During the data transmission phase, CSI is estimated from the orthogonal training sequences that are transmitted in the preamble of the data packets. With the estimated CSI, various MUD techniques can be applied to separate the multiple data packets at the AP. Using coherent detection, data packets can be transmitted at a much higher rate than the RTS packets without involving excessive computational complexity.

As MUD techniques have been introduced in Section II, we focus on the blind separation of RTS packets in this subsection. Assume that there are $K$ stations transmitting RTS packets together. Then, the received signal in symbol duration $n$ is given by (4), where the $(m, k)$th element of $\mathbf{H}$ denotes the channel coefficient from user $k$ to the $m$th antenna at the AP. Assuming that the channel is constant over an RTS packet, which is composed of $N$ symbol periods, we obtain the following block formulation of the data

$$\mathbf{Y} = \mathbf{HX} + \mathbf{W} \tag{41}$$

where $\mathbf{Y} = \left[\mathbf{y}(1), \mathbf{y}(2), \cdots \mathbf{y}(N)\right]$, $\mathbf{X} = \left[\mathbf{x}(1), \mathbf{x}(2), \cdots \mathbf{x}(N)\right]$, and $\mathbf{W} = \left[\mathbf{w}(1), \mathbf{w}(2), \cdots \mathbf{w}(N)\right]$. The problem to be addressed here is the estimation of the number of sources $K$, the channel matrix $\mathbf{H}$, and the symbol matrix $\mathbf{X}$, given the array output $\mathbf{Y}$.

*a) Estimation of the number of sources K*

For an easy start, we ignore the white noise for the moment and have $\mathbf{Y} = \mathbf{HX}$. The rank of **H** is equal to *K* if $K \le M$. Likewise, **X** is full-row-rank when *N* is much larger than *K*. Consequently, we have $rank(\mathbf{Y}) = K$ and *K* is equal to the number of nonzero singular values of **Y**. With white noise added to the data, *K* can be estimated from the number of singular values of **Y** that are significantly larger than zero.

*b) Estimation of* **X** *and* **H**

In this paper, we adopt the Finite Alphabet (FA) based blind detection algorithm to estimate **X** and **H**, assuming *K* is known. The maximum-likelihood estimator yields the following separable least-squares minimization problem [18]

$$\min_{\mathbf{H},\mathbf{X}\in\Omega} \left\| \mathbf{Y} - \mathbf{HX} \right\|_F^2 \tag{42}$$

where $\Omega$ is the finite alphabet to which the elements of **X** belong, and $\left\| \cdot \right\|_F^2$ is the Frobenius norm. The minimization of (42) can be carried out in two steps. First, we minimize (42) with respect to **H** and obtain

$$\hat{\mathbf{H}} = \mathbf{Y}\mathbf{X}^{+} = \mathbf{Y}\mathbf{X}^{H}\left(\mathbf{X}\mathbf{X}^{H}\right)^{-1}, \tag{43}$$

where $\left(\cdot\right)^{+}$ is the pseudo-inverse of a matrix. Substituting $\hat{\mathbf{H}}$ back into (42), we obtain a new criterion, which is a function of **X** only:

$$\min_{\mathbf{X}\in\Omega} \left\| \mathbf{Y}\mathbf{P}_{\mathbf{X}^H}^{\perp} \right\|_F^2, \tag{44}$$

where $\mathbf{P}_{\mathbf{X}^H}^{\perp} = \mathbf{I} - \mathbf{X}^{H}\left(\mathbf{X}\mathbf{X}^{H}\right)^{-1}\mathbf{X}$, and **I** is the identity matrix. The global minimum of (44) can be obtained by enumerating over all possible choices of **X**. Reduced-complexity iterative algorithms that solve (44) iteratively such as ILSP and ILSE were introduced in [19]. Not being one of the foci of this paper, the details of ILSP and ILSE are not covered here. Interested readers are referred to [20] and the references therein.

Note that the scheme proposed in this section is only one way of implementing MPR in WLANs. It ensures that the orthogonal training sequences are transmitted in the preambles of data packets. This leads to highly reliable channel estimation that facilitates the user of MUD techniques. In Appendix C, we briefly discuss another scheme that is applicable to both basic-access and RTS/CTS access modes.

## VI. CONCLUSION

With the recent advances in PHY-layer MUD techniques, it is no longer a physical constraint for the channel to accommodate only one ongoing packet transmission in WLANs. To fully utilize MPR capability of the PHY channel, it is essential to understand the fundamental impact of MPR on the MAC-layer. This paper has studied the characteristic behavior of random-access WLANs with MPR. Not only does our analysis lay down a theoretical foundation for the performance evaluation of WLANs with MPR, it is also a useful aid for system design in terms of setting operating parameters.

Our analytical framework is general and applies to various WLANs including non-carrier-sensing and carrier-sensing networks. In Theorem 1 and 3, we have proved that the throughput increases super-linearly with $M$ for both finite and infinite population cases. This is the case in non-carrier-sensing networks for all $M$, and in carrier-sensing networks for large $M$. Moreover, Theorem 2 shows that the throughput penalty due to distributed random access diminishes when $M$ approaches infinity. Such scalability provides strong incentives for further investigations on engineering and implementation details of MPR systems. Based on the analysis, we found that the commonly deployed BEB scheme is far from optimum in most systems except the carrier-sensing systems with RTS/CTS four-way handshake. In particular, the optimum $r$ increases with $M$ for large $M$. We further note that the throughput degrades sharply when $r$ is smaller than the optimum value, while the curve is much flatter when $r$ exceeds the optimum. Therefore, for system robustness, it is preferable to set $r$ large enough to avoid dramatic throughput degradation.

Having understood the fundamental behavior of MPR, we propose a practical protocol to exploit the advantage of MPR in IEEE 802.11-like WLANs. By incorporating advanced PHY-layer blind detection and MUD techniques, the protocol can implement MPR in a fully distributed manner with marginal modification of MAC layer.

## APPENDIX A: SUPER-LINEAR THROUGHPUT SCALING IN WLANS WITH FINITE POPULATION

*Theorem 3* (*Super-linearity with finite population*): $\dfrac{S_N^*(M+1)}{M+1} \geq \dfrac{S_N^*(M)}{M}$ for all $M < N$

From (12), we have

$$
\begin{aligned}
S_N(M, p_t) &= R\sum_{k=1}^{M} k\binom{N}{k} p_t^k(1-p_t)^{N-k} \\
&= R\frac{Np_t}{1-p_t}\sum_{k=0}^{M-1}\binom{N}{k} p_t^k(1-p_t)^{N-k} - R\frac{p_t}{1-p_t}\sum_{k=0}^{M-1} k\binom{N}{k} p_t^k(1-p_t)^{N-k} \\
&= R\frac{Np_t}{1-p_t}\Pr\left\{X \le M-1\right\} - \frac{p_t}{1-p_t}S_N(M-1, p_t) \text{ bits/sec}
\end{aligned} \tag{A1}
$$

and
$$
S_N(M+1, p_t) = R\frac{Np_t}{1-p_t}\Pr\left\{X \le M\right\} - \frac{p_t}{1-p_t}S_N(M, p_t). \tag{A2}
$$

Meanwhile,
$$
S_N(M+1, p_t) = R\sum_{k=1}^{M+1} k\binom{N}{k} p_t^k(1-p_t)^{N-k} = S_N(M, p_t) + R(M+1)\Pr\left\{X = M+1\right\} \tag{A3}
$$

Substituting (A3) to (A2), we get

$$
S_N(M, p_t) = RNp_t\Pr\left\{X \le M\right\} - R\left(1-p_t\right)(M+1)\Pr\left\{X = M+1\right\} \quad \forall M < N, p_t. \tag{A4}
$$

At the optimal $p_t^*(M)$, the derivative $\partial S_N(M, p_t)/\partial p_t = 0$. Thus,

$$
\begin{aligned}
&\left.\frac{\partial S_N(M, p_t)}{\partial p_t}\right|_{p_t = p_t^*(M)} = \\
&RN\Pr\left\{X \le M\right\}\Big|_{p_t = p_t^*(M)} + R(M+1)\left(1-(M+1)\frac{1}{p_t}\right)\Pr\left\{X = M+1\right\}\Big|_{p_t = p_t^*(M)} = 0
\end{aligned}, \tag{A5}
$$

Combining (A4) and (A5),

$$
\begin{aligned}
&S_N(M, p_t^*(M)) = p_t^*(M)\left.\frac{\partial S_N(M, p_t)}{\partial p_t}\right|_{p_t = p_t^*(M)} \\
&-R(M+1)\left(p_t^*(M)-(M+1)\right)\Pr\left\{X = M+1\right\}\Big|_{p_t = p_t^*(M)} - R(M+1)\left(1-p_t^*(M)\right)\Pr\left\{X = M+1\right\} \\
&= RM(M+1)\Pr\left\{X = M+1\right\}\Big|_{p_t = p_t^*(M)}
\end{aligned} \tag{A6}
$$

It is obvious that

$$
S_N(M+1, p_t^*(M+1)) \ge S_N(M+1, p_t^*(M)) = S_N(M, p_t^*(M)) + R(M+1)\Pr\left\{X = M+1\right\}\Big|_{p_t^*(M)} \tag{A7}
$$

Substituting (A6) to (A7), we have

$$
S_N(M+1, p_t^*(M+1)) \ge S_N(M, p_t^*(M)) + \frac{S_N(M, p_t^*(M))}{M} = S_N(M, p_t^*(M))\frac{M+1}{M}. \tag{A8}
$$

Hence, $\frac{S_N^*(M+1)}{M+1} \ge \frac{S_N^*(M)}{M} \quad \forall M < N$. □

APPENDIX B Proof of Lemma 1 (b) and 1(c)

$$\begin{aligned} S_\infty(M) &= R\lambda \Pr\left\{X \le M-1\right\} = R\lambda \sum_{k=0}^{M-1} \frac{\lambda^k}{k!} e^{-\lambda} \\ &\le R\lambda z^{-M} \sum_{k=0}^{M-1} \frac{(\lambda z)^k}{k!} e^{-\lambda} \le R\lambda z^{-M} \sum_{k=0}^{\infty} \frac{(\lambda z)^k}{k!} e^{-\lambda} = R\lambda z^{-M} e^{\lambda(z-1)} \quad \forall z<1 \end{aligned} \tag{B1}$$

Let $g(z) = R\lambda z^{-M} e^{\lambda(z-1)}$ be the upper bound of $S_\infty(M)$. By solving

$$\frac{\partial g(z)}{\partial z} = R\lambda\left(-Mz^{-M-1}e^{\lambda(z-1)} + \lambda z^{-M} e^{\lambda(z-1)}\right) = 0 \tag{B2}$$

it can be easily found that $z^* = M/\lambda$ minimizes $g(z)$ and

$$\frac{g(z^*)}{\lambda R} = \left(\frac{\lambda}{M}\right)^M e^{M\left(1-\frac{\lambda}{M}\right)}. \tag{B3}$$

Since $z^* < 1$, $\lambda > M$. Let $\lambda = cM$ where $c > 1$. Eqn. (B3) can be written as

$$\frac{g(z^*)}{\lambda R} = \left(ce^{(1-c)}\right)^M = \left(ce^{(1-c)}\right)^M. \tag{B4}$$

Due to Eqn. (23)

$$\lim_{M\to\infty} \frac{S_\infty(M)}{\lambda R} \le \lim_{M\to\infty} \frac{g^*(z)}{\lambda R} = \lim_{M\to\infty} \left(ce^{(1-c)}\right)^M = 0 \tag{B5}$$

On the other hand, it is obvious that

$$\frac{S_\infty(M)}{\lambda R} \ge 0 \tag{B6}$$

Combining (B5) and (B6), we have

$$\lim_{M\to\infty} \frac{S_\infty(M)}{\lambda R} = 0 \;\; \forall \lambda > M, \tag{B7}$$

and Lemma 1(b) follows.

To prove Lemma 1(c), we note that the median of Poisson distribution is bounded as follows [21]-[22]:

$$\lambda - \log 2 \le \text{median} \le \lambda + 1/3. \tag{B8}$$

When $\lambda = M$ and $M \to \infty$, the median approaches *M*. According to the first equality of (19),

$$\lim_{M\to\infty} \frac{S_\infty(M)}{\lambda R} = \lim_{M\to\infty} \Pr\left\{X \le M-1\right\} = \lim_{M\to\infty} \Pr\left\{X \le M\right\} = 0.5. \qquad \square$$

## APPENDIX C MPR IN IEEE 802.11 BASIC ACCESS MODE

In Section VI, we have introduced a MAC-PHY protocol for implementing MPR in IEEE 802.11 RTS/CTS WLANs. The protocol relies on the blind separation of low-rate RTS packets to identify senders and allocate orthogonal training sequences to them. In this section, we briefly describe an alternative protocol that is applicable to both basic-access mode.

The protocol works as follows. Upon transmitting a data packet, a station randomly picks a training sequence from a pool of orthogonal sequences. If more than one station picks the same sequence, then the receiver cannot resolve these transmissions and a collision happens between the packets involved. For other packets that have selected a unique orthogonal sequence, the receiver can estimate their CSI, and consequently apply MUD schemes to separate the packets.

Note that the number of orthogonal sequences is proportional to the length of the training sequences. On one hand, a small number of orthogonal sequences lead to a higher probability of collision. On the other hand, having more sequences leads to a larger overhead for channel estimation. Therefore, the number of sequences is a parameter to be optimized in this protocol.

The proposed protocols have their respective disadvantages and advantages. The one proposed in this section requires a lower computational complexity, as there is no need to do blind detection. However, collision may happen even if less than $M$ stations attempt to transmit together, when more than one station picks the same sequence. In contrast, the one proposed in Section VI guarantees that packets can be resolved as long as the number of attempts is smaller than $M$ (assuming small noise). It also ensures that there is no collision between training sequences of data packets. However, the computational complexity could be high if the RTS packets are transmitted at a high rate.

TABLE I

SYSTEM PARAMETERS USED IN CARRIER-SENSING NETWORKS (ADOPTED FROM IEEE 802.11G)

| | |
|---|---|
| Packet payload | 8184 bits |
| MAC header | 272 bits |
| PHY overhead | 26 μs |
| ACK | 112 bits + PHY overhead |
| RTS | 160 bits + PHY overhead |
| CTS | 112 bits + PHY overhead |
| Basic rate | 6 Mbps |
| Data rate | 54 Mbps |
| Slot time $\sigma$ | 9 μs |
| SIFS | 10 μs |

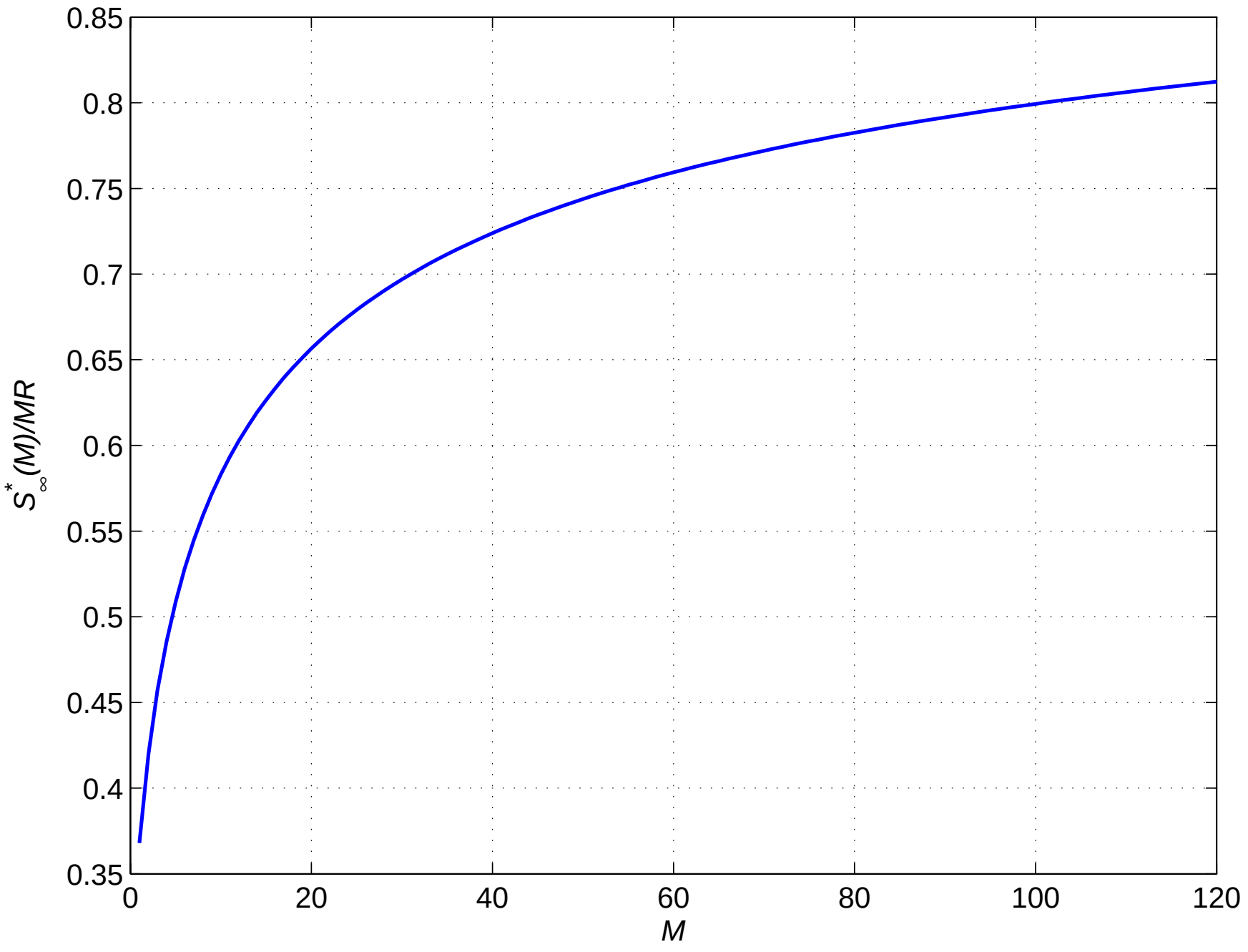

**Fig. 1:** Super-linear scalability of the throughput of non-carrier sensing slotted ALOHA networks

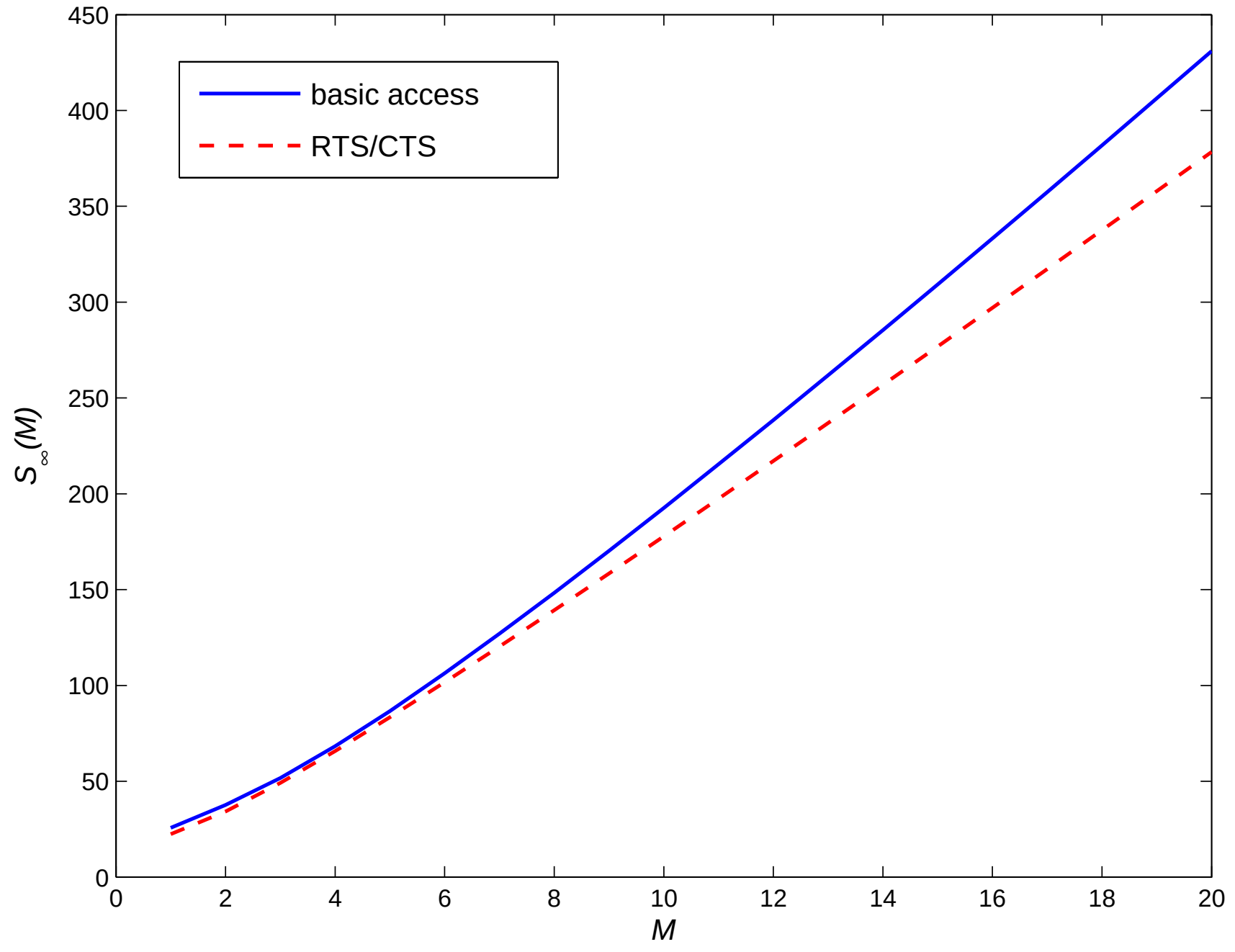

**Fig. 2** Optimal throughput of carrier-sensing networks

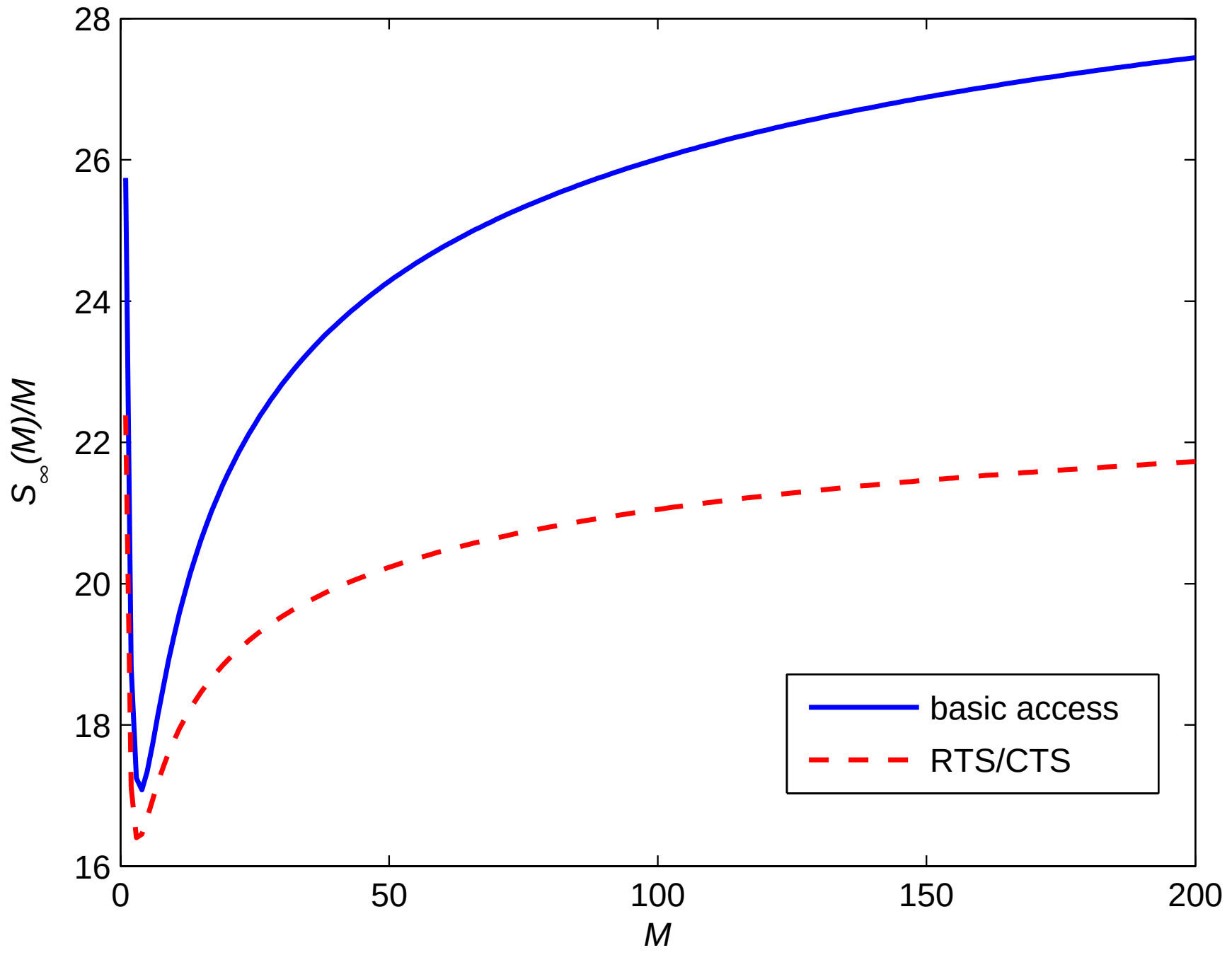

**Fig. 3:** Super-linear scalability of the throughput of carrier-sensing networks

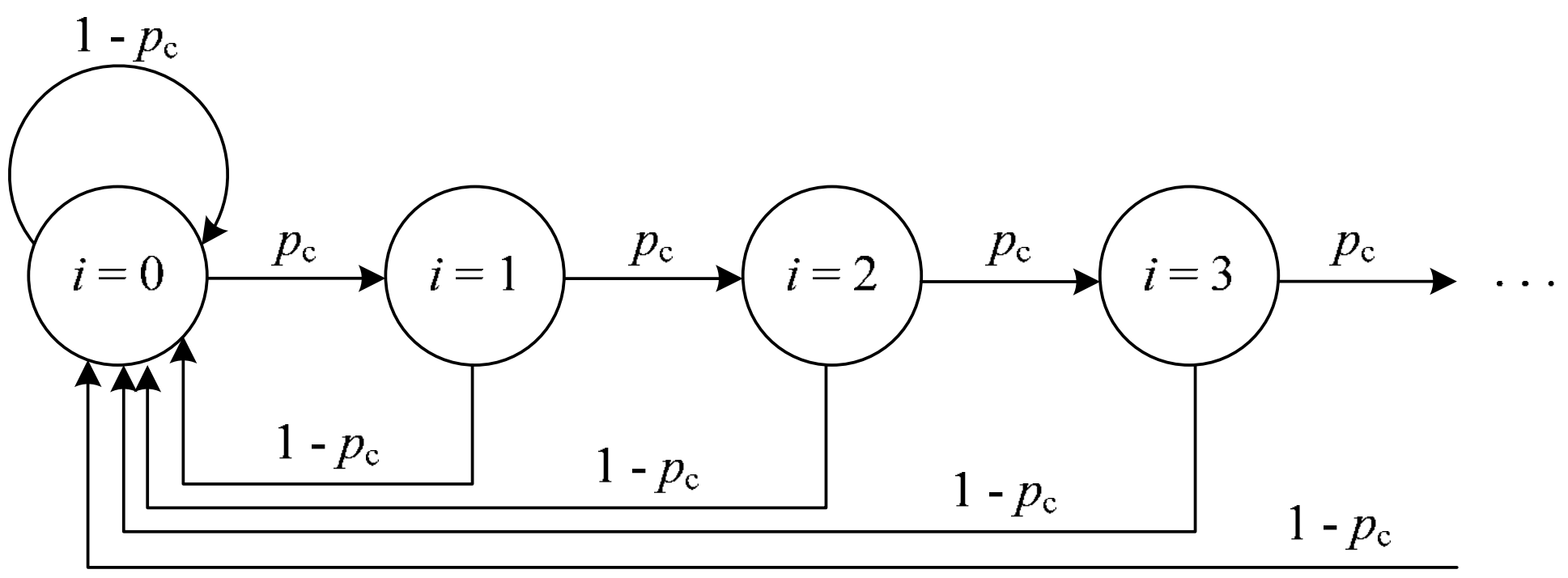

**Fig. 4:** Markov chain model for the backoff stage.

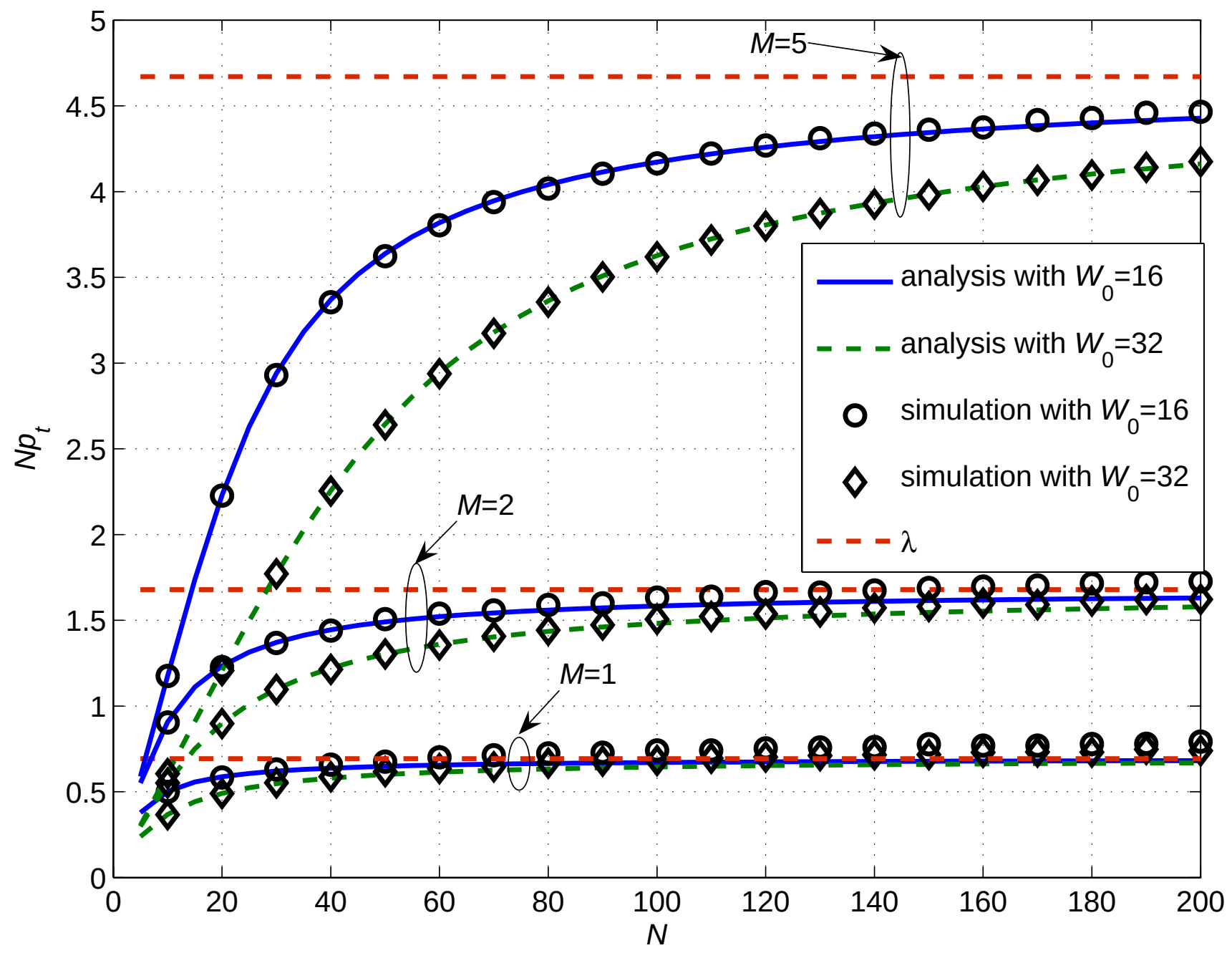

**Fig. 5:** Plots of $Np_t$ versus $N$ when $r$ = 2; lines are analytical results calculated from (2) and (3), markers are simulation results.

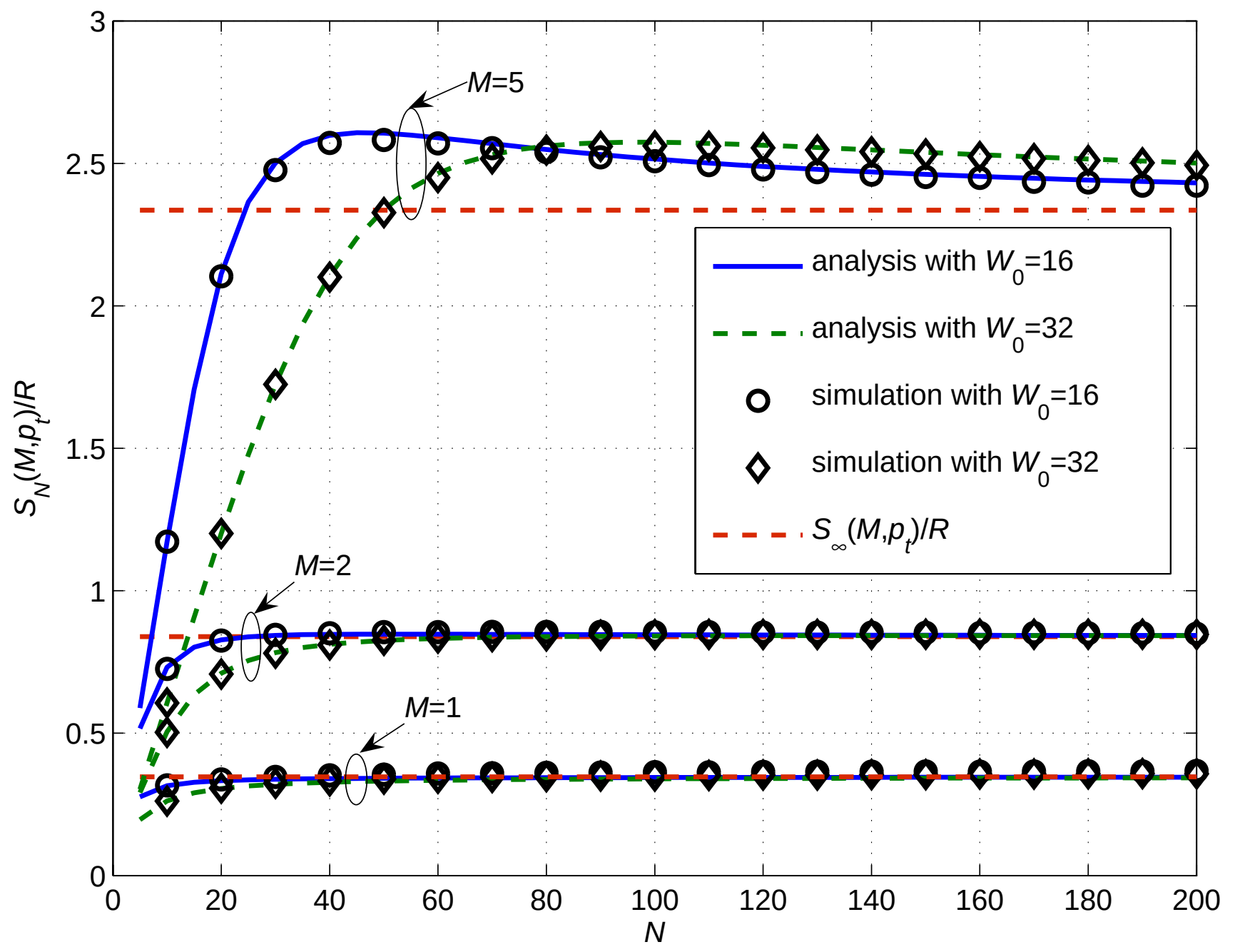

**Fig. 6**: Normalized throughput of non-carrier-sensing slotted ALOHA networks when $r = 2$

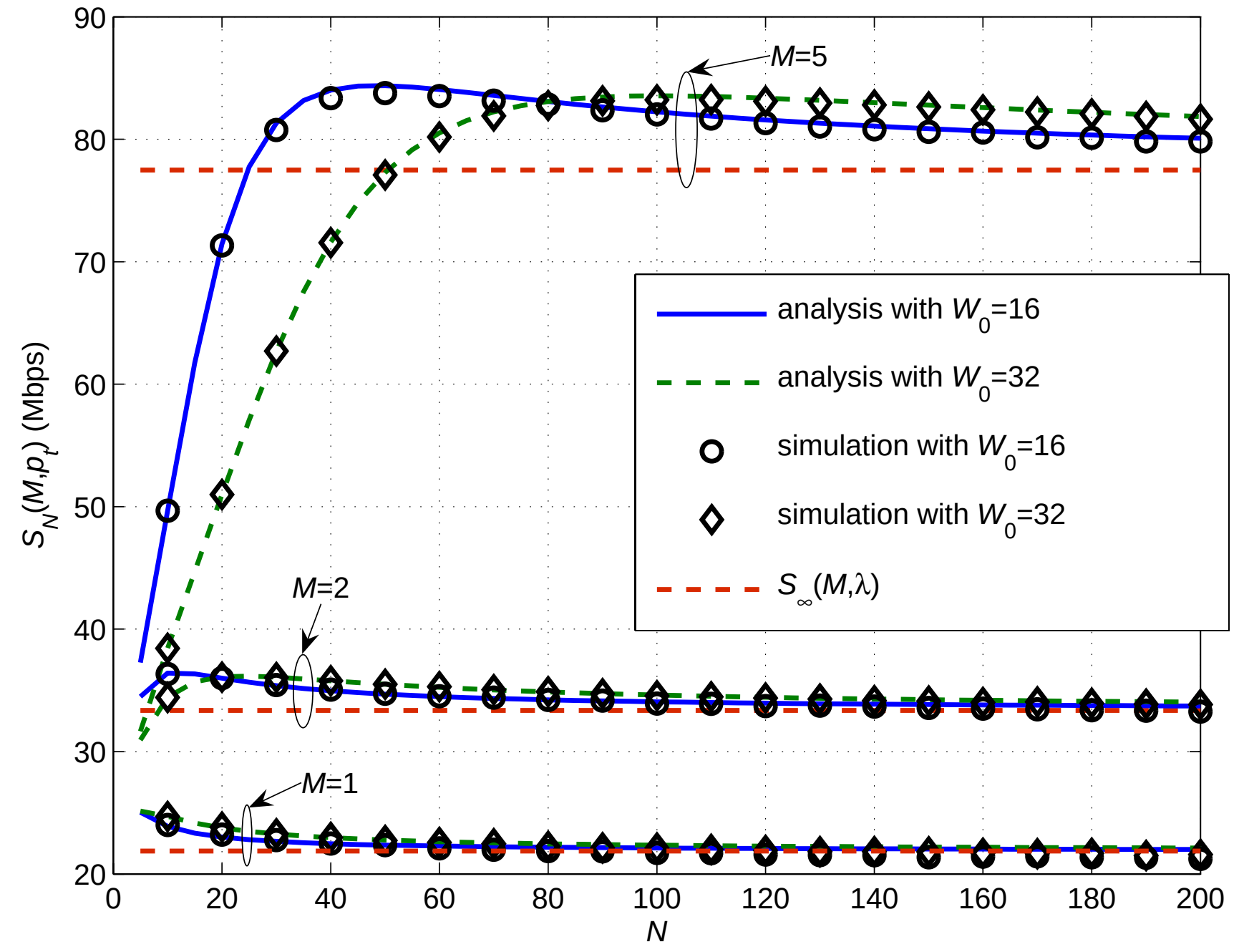

**Fig. 7**: Throughput of carrier-sensing basic-access networks when $r = 2$

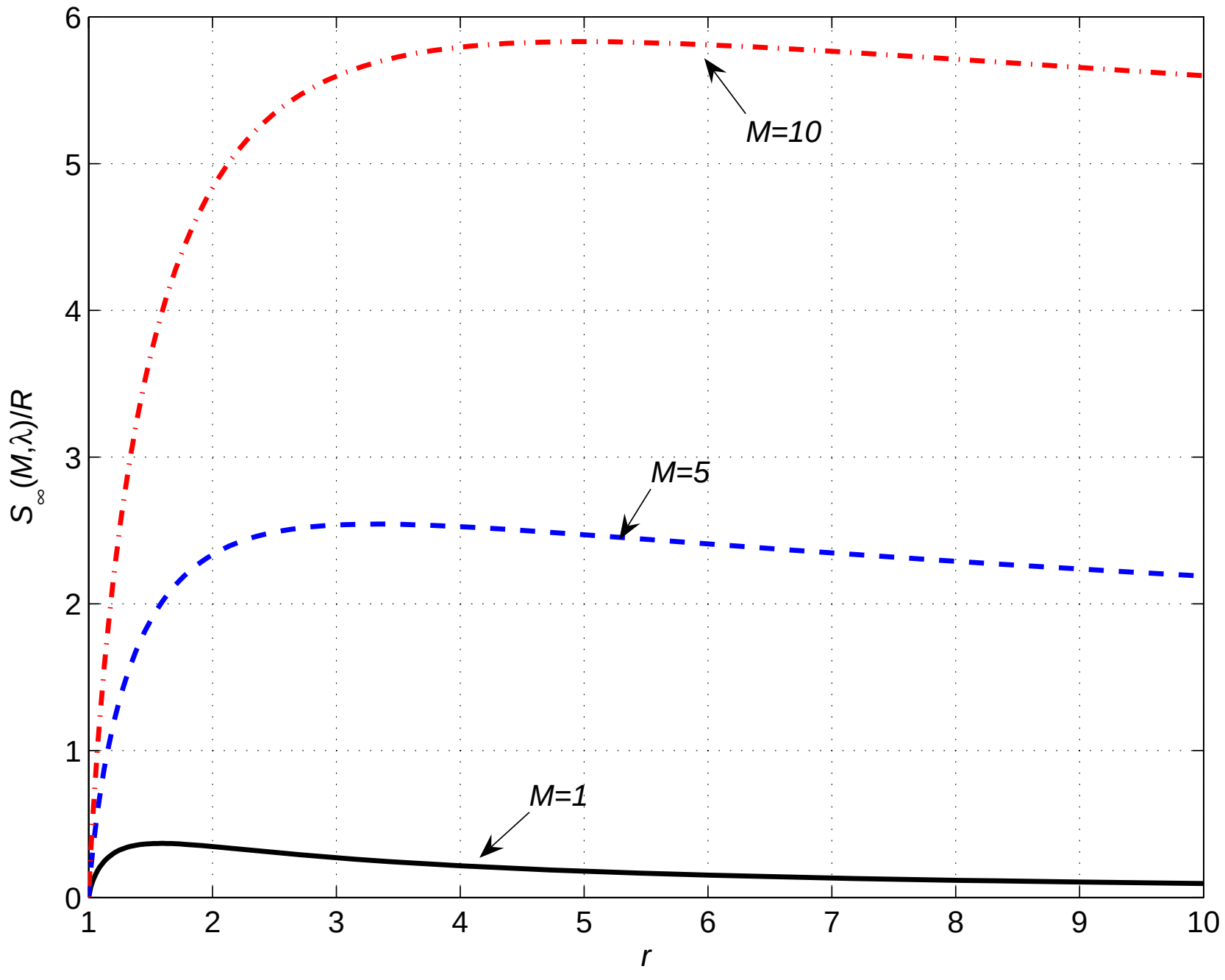

**Fig. 8:** Throughput versus r for non-carrier-sensing slotted ALOHA networks

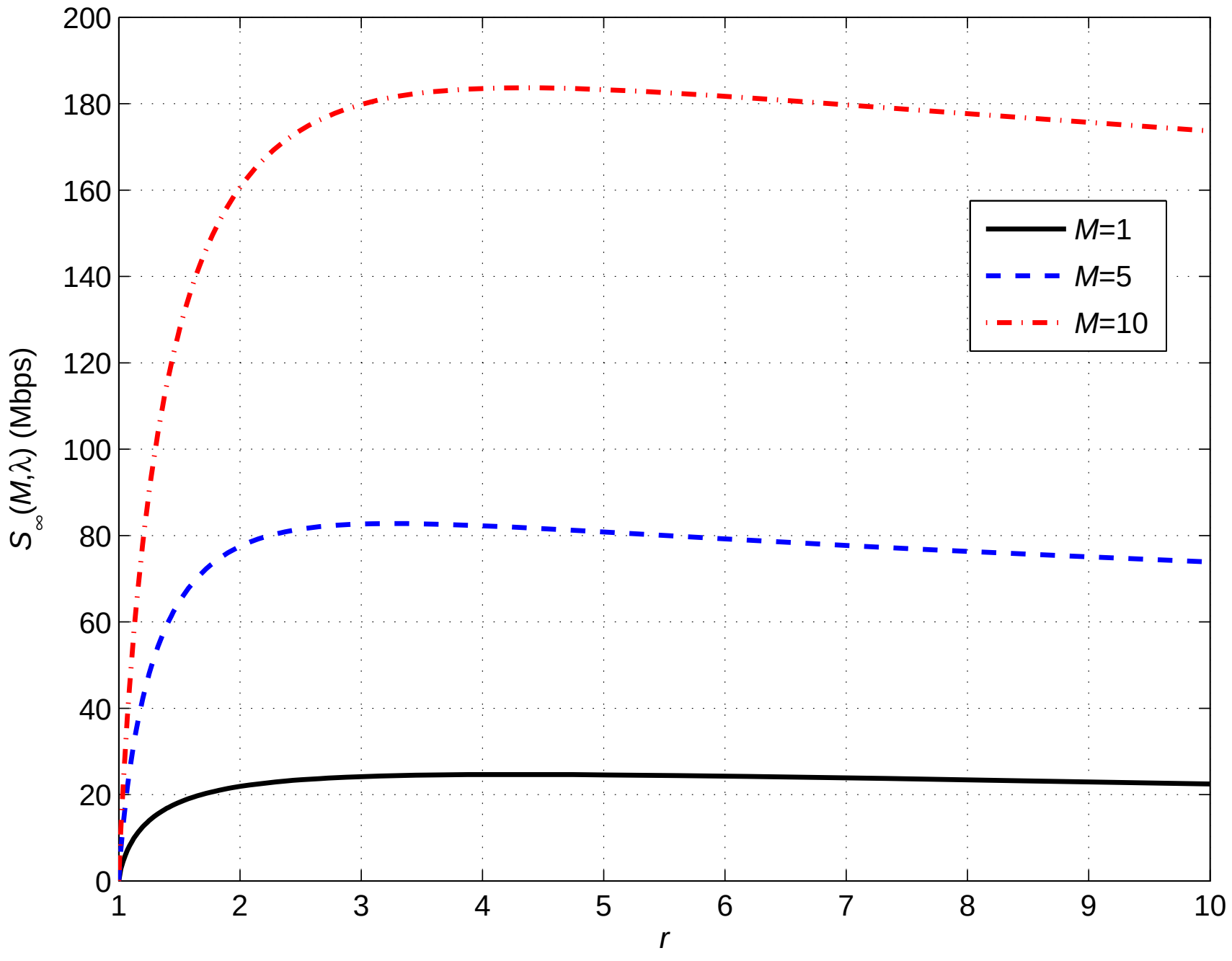

**Fig. 9:** Throughput versus *r* for carrier-sensing networks with basic-access mode

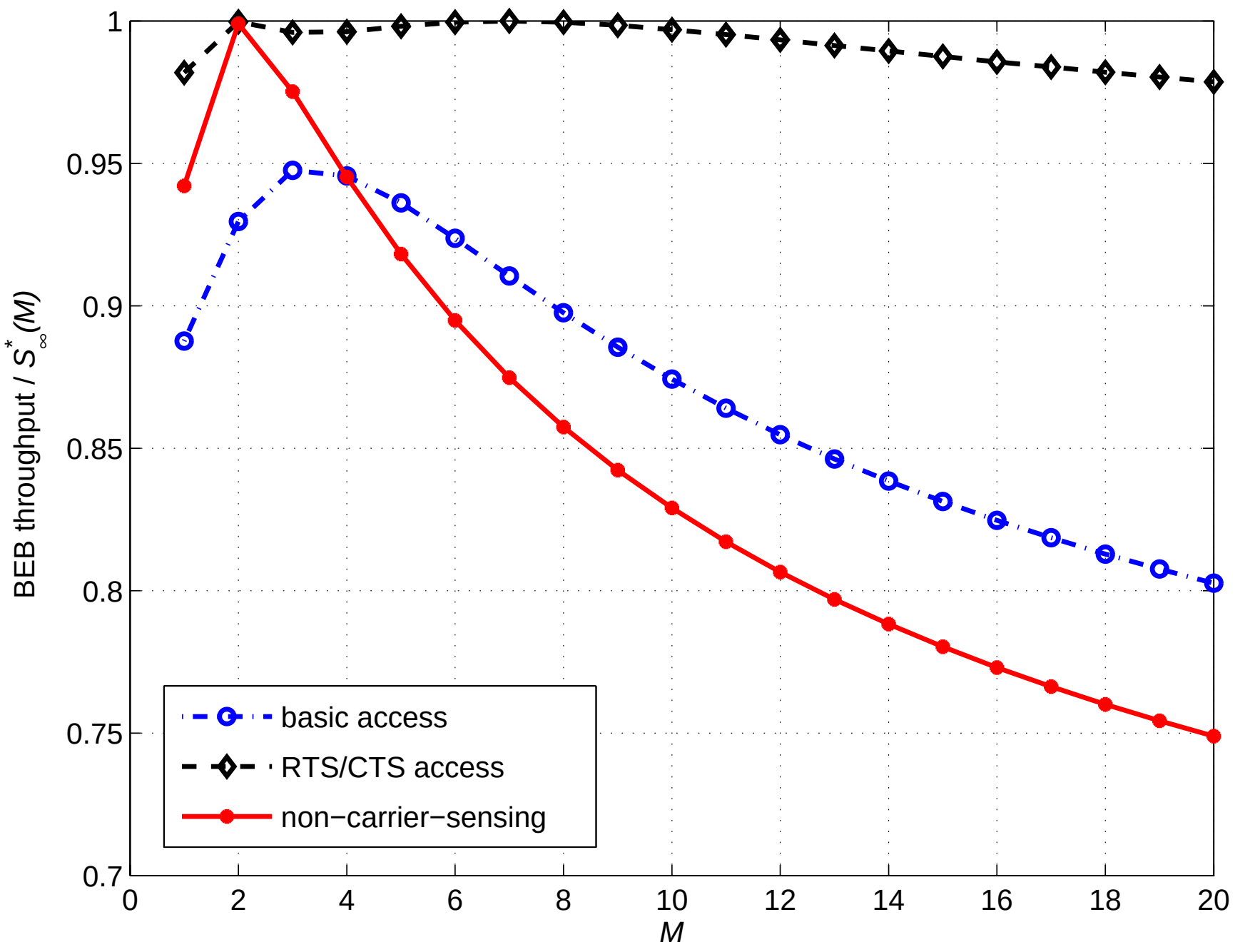

**Fig. 10**: Ratio of the throughput of BEB to maximal throughput versus *M*

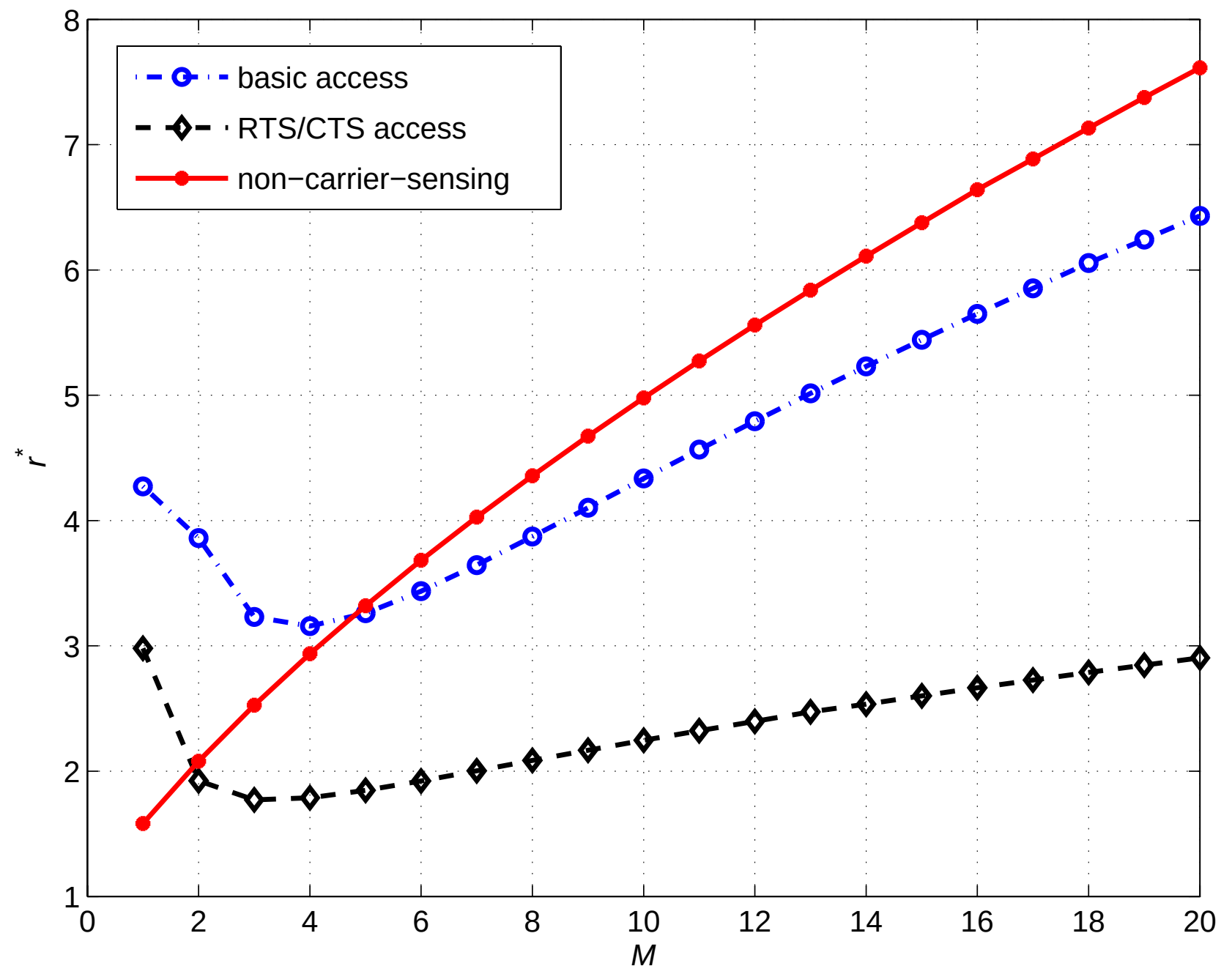

**Fig. 11**: Optimal *r* versus *M*

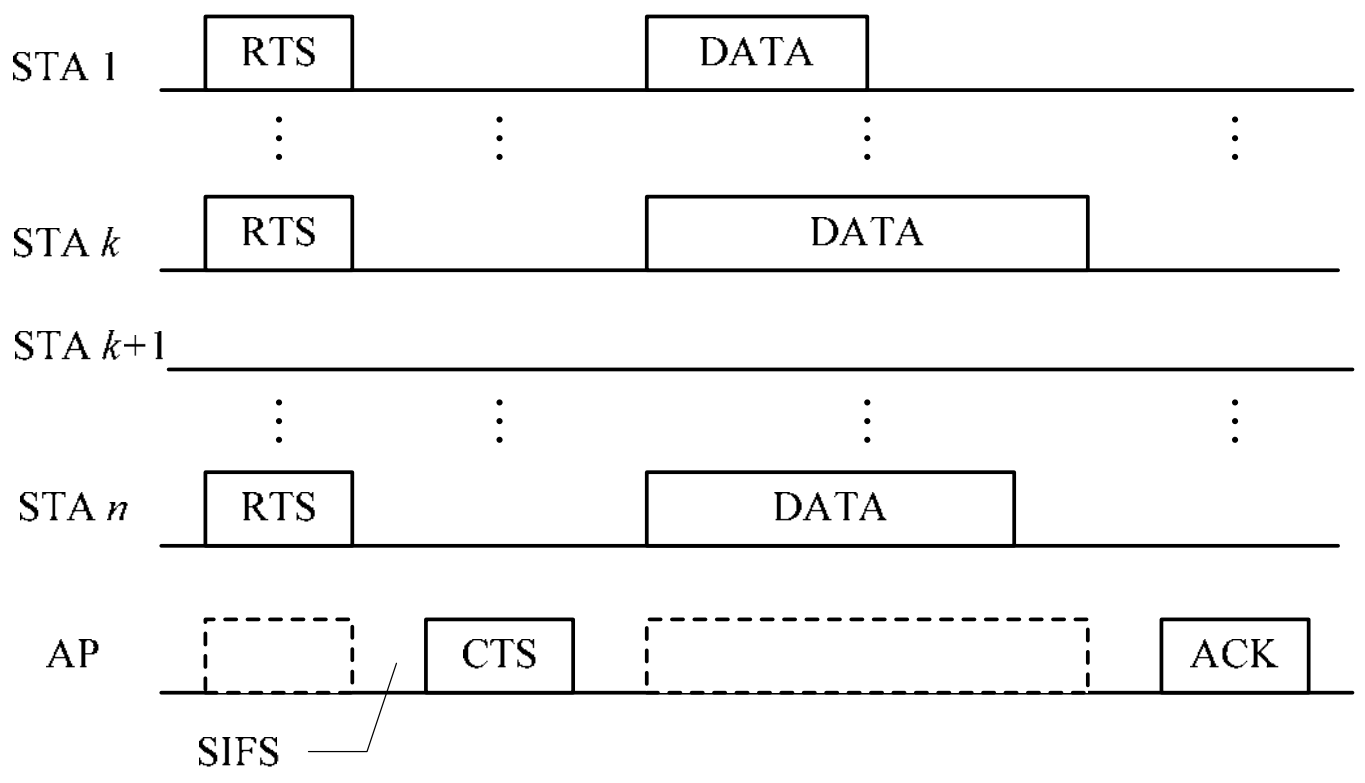

**Fig. 12:** Time line example for the MPR MAC